# Measuring Domestic Violence. Individual Attitudes and Time Use Within the Household

Elena Pisanelli
University of Bergamo, Department of Economics

## Abstract

This paper proposes a novel empirical strategy to measure cultural justifications of domestic violence within households, with direct implications for demographic behavior and gender inequality. Leveraging survey data on individual attitudes and high-frequency time-use diaries from Italian couples with children, I construct a composite index that integrates stated beliefs with observed household practices. Using structural equation modeling, I disentangle latent tolerance of domestic violence from reported attitudes and validate the index against both individual and partner characteristics, as well as time allocation patterns. Results reveal systematic heterogeneity by gender, education, and normative environments. Conservative gender and parenthood norms are strong predictors of tolerance, while higher male education reduces it. Tolerance of violence is also positively associated with reported leisure time with partners and children, suggesting that co-presence does not necessarily reflect egalitarian interaction but may coexist with unequal bargaining structures. Beyond advancing measurement, the findings highlight how cultural tolerance of domestic violence is embedded in household arrangements that influence fertility, labor supply, and the intergenerational transmission of norms. The proposed framework offers a scalable tool for economists and policymakers to monitor hidden inequalities and design interventions targeting family stability, gender equity, and child well-being.

The research is funded by the European Union - Next Generation EU, in the framework of the GRINS - Growing Resilient, Inclusive and Sustainable Project (GRINSPE00000018 – CUP J33C22002910001). The views and opinions expressed are solely those of the authors and do not necessarily reflect those of the European Union, and the European Union cannot be held responsible for them.

# 1. Introduction

Understanding domestic violence is central to economics because of its wide-ranging consequences for family formation, fertility decisions, labor market participation, and intergenerational well-being. Intimate partner violence not only undermines individual health and safety, but also constrains women's labor supply, shapes bargaining power within households, and influences investments in children. Despite decades of research, one of the main challenges for both scholars and policymakers remains measurement. How can we capture attitudes and norms that legitimize violence when such beliefs are often unreported, misrepresented, or internalized as culturally acceptable?

Traditional survey-based approaches, while indispensable, tend to underestimate prevalence due to stigma and social desirability bias. As a result, blind spots remain in understanding how norms of control and coercion shape household dynamics. This limits our ability to assess the demographic consequences of violence—on fertility, household stability, and gendered labor allocation—and to design effective policy responses.

This paper addresses these challenges by proposing a new measurement strategy that integrates attitudinal data with high-frequency behavioral data from household time-use diaries. Drawing on original data from the TIMES project in Italy, I develop an empirical framework that links beliefs about gender roles and violence with observed behavior—specifically, how couples allocate time between partners and children. By combining vignette-based survey responses with diary-based records of unpaid work and leisure, I construct a composite indicator that captures latent cultural orientations toward domestic violence.

The contribution of this paper is threefold. First, methodologically, it develops a novel indicator that blends direct and indirect measures of domestic violence tolerance through structural equation modeling. Second, substantively, it shows that attitudes justifying violence are systematically related to gendered household structures and time allocation, shedding light on mechanisms of inequality within families. Third, it shows that tolerance of violence is not merely an individual belief but a household-level feature with implications for fertility behavior, women's labor supply, and the intergenerational transmission of gender norms. Importantly, results show that tolerant attitudes are associated not with withdrawal from joint activities but with more reported leisure time with partners and children, underscoring that co-presence may mask persistent asymmetries in bargaining power.

By linking domestic violence attitudes to observable household practices, this study expands the toolkit of economists, offering a scalable approach to measuring hidden dimensions of inequality. The framework has broad relevance for contexts where direct reporting of violence is unreliable and where understanding the cultural underpinnings of household behavior is crucial for explaining demographic outcomes and informing policy.

# 2. Literature Review

Research on domestic violence has grown considerably in economics, with increasing attention to its measurement, economic determinants, and interaction with gender norms. Domestic violence is not only a matter of individual welfare but also a factor influencing household formation, fertility, labor supply, and intergenerational inequality. This section situates my contribution within three main strands of literature: (i) methodological advances in measurement, (ii) economic and demographic determinants of violence, and (iii) the role of gender norms and attitudes in shaping household outcomes.

## 2.1 Measurement Challenges and Methodological Innovations

A persistent challenge in the study of domestic violence is underreporting. Fear of retaliation, stigma, and social desirability biases often lead victims to conceal experiences, especially in household surveys (Cullen, 2023). To address this, researchers have adopted indirect methods such as list experiments or randomized response techniques, which sometimes yield substantially higher prevalence rates (Cullen, 2023; Agüero & Frisancho, 2022). Other studies have turned to community reporting or administrative data, though these often underestimate true incidence.

Recent work has also questioned the validity of standard survey instruments, such as the Conflict Tactics Scale, for capturing the complexity of coercive control and psychological abuse (Clark et al., 2024). To supplement surveys, alternative behavioral data sources have been explored. For example, Anderberg et al. (2020) use Google search trends during the COVID-19 lockdowns to show that spikes in domestic violence-related queries tracked emergency hotline calls more closely than police reports, suggesting official statistics severely understate actual incidence.

These methodological innovations underscore the need to move beyond self-reports and incorporate behavioral indicators. My paper contributes by leveraging time-use diaries—an underutilized but promising data source for understanding relational dynamics. By linking attitudinal items with household time allocation, I provide an integrated framework less vulnerable to reporting bias, while also embedding domestic violence within the broader study of household economics.

## 2.2 Economic and Demographic Determinants of Domestic Violence

From a household economics perspective, domestic violence is closely tied to bargaining models and intra-household allocation. Early theoretical contributions view violence as a strategic instrument to assert control within households (Anderberg et al., 2016). Empirically, shocks to male and female employment have asymmetric effects: male job loss tends to reduce domestic violence, while female job loss increases vulnerability (Anderberg et al., 2016). This is consistent with models where bargaining power is shaped by outside options.

The demographic literature has highlighted how exposure to economic shocks can alter violence patterns. During the COVID-19 pandemic, job losses in Peru were associated with sharp increases in intimate partner violence (Agüero et al., 2024). Similarly, Carr and Packham (2019) show that changes in U.S. SNAP disbursement schedules increased violence, highlighting how policy design affects household conflict. These findings stress that violence is not only a private phenomenon but also a population-level outcome influenced by labor markets, social insurance, and economic institutions.

Violence also has long-term demographic consequences. It can affect women's labor force participation, fertility choices, and investments in children. Showalter (2016) documents how women's employment reduces vulnerability but may heighten tensions in contexts with rigid gender norms. Tur-Prats (2019) shows that historical family structures in Spain influenced current violence prevalence and women's autonomy, suggesting intergenerational persistence of norms. Together, these studies illustrate that domestic violence is central to understanding household behavior, fertility, and intergenerational inequality.

Recent contributions also highlight the central role of stereotypes and household bargaining in shaping both perceptions and experiences of violence. Grembi, Rosso, and Barili (2024) construct an individual-level measure of gender stereotypes and domestic violence perceptions in Italy, showing that stronger stereotypes are linked to downplaying severity and greater victim-blaming. Similarly, Zhang (2023) examines the "female breadwinning paradox" in Australia, finding that when women earn more than their male partners, the risk of violence rises sharply, including a 33% increase in physical abuse and a 20% increase in emotional abuse. These studies underscore how deviations from traditional gender roles can provoke backlash and intensify conflict within households. My paper complements this literature by moving beyond stated perceptions or income-based deviations, linking attitudes with time-use behavior to construct a validated composite indicator of tolerance for domestic violence.

## 2.3 Gender Norms, Attitudes, and the Justification of Violence

Cultural norms shape both the prevalence and acceptability of domestic violence. In many contexts, a significant share of the population justifies intimate partner violence under specific conditions, with acceptance strongly correlated with lower education, poverty, and patriarchal structures (Wang, 2016; Huang et al., 2024). Such attitudes influence not only immediate risk but also broader household dynamics, including fertility decisions, division of labor, and child-rearing practices.

Economic studies emphasize how changes in women's economic opportunities affect violence, depending on normative environments. Frankenthal (2023) shows that increases in women's agricultural productivity in Peru reduced physical abuse, especially where patriarchal norms were stronger. Relatedly, Reitmann et al. (2020) demonstrate that framing and priming can shift attitudes toward violence, indicating that cultural tolerance is malleable.

These insights align with research showing that norms affect fertility and labor supply decisions, often through intra-household bargaining mechanisms. By linking justification of violence to time-use inequalities and household leisure, my paper captures how beliefs about masculinity and gender roles manifest in everyday household structures. This contributes to the literature by providing a composite, behaviorally grounded measure of norms that influence demographic outcomes. Existing research often assumes that tolerance of violence reduces relational cohesion. However, my results suggest the opposite correlation: tolerant attitudes are linked to greater reported joint leisure. This indicates that co-presence should not be interpreted as a straightforward marker of equality but rather may reflect traditional authority structures in which men define or dominate shared time.

## 2.4 Contributions and Gaps

Despite progress, several gaps remain. First, the literature has focused largely on physical violence, with less attention to psychological abuse and coercive control. Second, studies often separate attitudinal surveys from behavioral measures of household dynamics, missing the intersection of beliefs and daily practices. Third, few contributions explicitly link domestic violence measurement to economic questions such as fertility, household stability, or intergenerational transmission of norms.

My paper addresses these gaps by introducing a novel composite indicator that integrates attitudinal and behavioral data. Methodologically, this approach reduces bias from self-reporting. Substantively, it embeds tolerance of violence within household structures, showing how norms translate into inequalities in unpaid work and leisure. The contribution is to provide a scalable tool for identifying hidden cultural drivers of demographic behavior, with implications for fertility, labor supply, and the persistence of gender inequality across generations.

## 3. Data and Methodology

The data used in this study is part of the TIMES project, developed by researchers at the University of Bologna.

Data collection followed a two-stage process targeting individuals residing in Emilia Romagna and Campania, living in cohabiting couples with at least one child under the age of 11. A total of 2,286 individuals were recruited, with participation being individual and voluntary for each partner. This study relies, however, on matched survey data collected from 848 couples –1,696 individuals– with both partners independently reporting on their time use, gender norms, and attitudes toward intimate partner violence. Eligibility was strictly conditioned on the presence of at least one child living in the household, in order to focus on families facing early shared responsibilities, such as childcare. A unique feature of the data is that it captures responses from both partners within the same household, providing valuable insights into household dynamics.

The sample is stratified at the provincial level, and the sampling strategy includes quotas for gender, individual occupational status, and the size of the municipality of residence (0–10k, 10k–50k, >50k inhabitants). Additionally, participants were recruited proportionally across provinces to reflect the age structure of the national infant population (0–2, 3–5, and 6–10 years). After data validation, the interview weighting was found to be 97.6% efficient, indicating that the application of sampling weights—used to adjust for discrepancies between the sample and the target population—led to only a negligible loss in statistical efficiency.

Participants first completed a socio-economic questionnaire administered online via a custom-designed web app (CAWI methodology), accessible from computers, tablets, or smartphones, and available in both Italian and English. Approximately two weeks after completing the questionnaire, participants were invited to fill out a time-use diary covering two full days: one weekday and one weekend day. For an interview to be considered valid, participants had to complete both the questionnaire and the two diaries. The survey recorded a dropout rate of 2.07%, corresponding to participants who dropped out after completing the questionnaire but before starting the diaries.

**Questionnaire**. The questionnaire combines established measures—such as gender norm items from the World Values Survey (Inglehart, 2004)—with original items specifically developed for this study. It assesses beliefs about family management, intimate partner violence (IPV), masculinity, and social norms, using a 0–100 continuous scale for most attitudinal items.

**Time-Use Data**. The TIMES project collected high-frequency data on individuals' daily activities through digital time-use diaries. Digital diaries are increasingly adopted for their convenience, structured classification of tasks, and reduced coding error (Minnen et al., 2014; Bigoni et al., 2023). Each respondent completed two diaries: one referring to a randomly assigned weekday and one to a randomly assigned weekend day. This dual-day design enables the construction of weekly estimates of time allocated to different types of activities, using the formula:

$$Average\,Weekly\,Hours = 5 \times Weekday\,Hours + 2 \times Weekend\,Hours$$

Participants recorded all activities across the full 24-hour span of each assigned day using a web interface. Activities were selected from a pre-determined hierarchical list and recorded in 10-minute intervals, specifying the primary and (when applicable) secondary activity, the presence of others, and engagement with children. This method, based on the structure proposed by Bigoni et al. (2023), minimizes recall bias and allows for fine-grained behavioral analysis. Seasonal effects were mitigated by collecting data uniformly across the calendar year.

These data also allow the construction of broader behavioral indicators from the items relevant to gender dynamics, violence, and parenting—such as time spent with one's partner or with children.

## 3.1 Survey Items and Indicator Construction

To enhance clarity, this section introduces the survey items first, followed by how they are grouped into indicators, and finally how these indicators contribute to the latent variables used in the SEM.

### 3.1.1 Item-to-Indicator Structure

The survey includes multiple items capturing attitudes toward domestic violence, gender roles, and household labor. These items are used to construct observed indicators for three latent constructs: justification of domestic violence, masculinity norms, and gender gap in unpaid work.

#### 3.1.1.1 Justification of Domestic Violence

The key variables used to construct justification of domestic violence indicator are drawn from a combination of a vignette-based design and survey questions. Each respondent was randomly assigned one of two hypothetical scenarios:

Scenario 1:

"*Sara and Davide have been a couple for 10 years. During one of their many arguments, Sara started yelling and Davide slapped and hit her.*"

Scenario 2:

"*Sara and Davide have been a couple for 10 years. When Sara goes out at night, Davide constantly messages her to ask what she is doing, where she is, and whom she is with*."

After viewing one of the vignettes, respondents rated their agreement with the following statements on a 0–100 scale:

- Seriousness of Violence: "The scenario described is serious."

- Victim Blaming: "Sara is responsible for Davide's behavior."

- Perpetrator Accountability: "Davide is responsible for his behavior."

- Justification of Domestic Violence: "Violence against women/men is justified."

#### 3.1.1.2 Masculinity Norms

Items capturing the endorsement of traditional masculine norms. Respondents rated their agreement with the following statements on a 0–100 scale:

- Minimization of Harassment: "Too much nonsense is spoken about socalled sexual harassment."

- Problematic Masculinity Traits: Agreement with the statements "It is not acceptable for a man to cry." (Emotional strength); "Drinking heavily is not a sign of masculinity but a problem." (Drinking) "Physical strength is a fundamental aspect of being a man" (Physical strength) and "Sensitivity is an admirable trait for all genders." (Emotional toughness).

### 3.1.1.3 Gender gap in unpaid work

Time-use variables, coming from time-use diaries, calculated as the relative difference in time spent by women and men in housework and childcare, computed as (female − male)/male.

Gender gap in household chores: The difference between female and male time spent on activities such as meal preparation and clean-up, doing laundry, ironing, dusting, vacuuming, indoor cleaning, constructing or repairing household items, purchasing goods or services for the family, and managing family life (e.g., planning visits, budgeting). The difference is weighted by the male time spent on the same activities.

Gender gap in childcare: The difference between female and male time spent on childcare activities, including putting the child to bed or waking them up; helping with eating, bathing, dressing, or grooming; reading; listening to the child read; teaching (reading, writing, counting); playing; watching cartoons; visiting museums, exhibitions, theaters, or zoos; doing artistic, manual, or creative activities; watching television, films, or series; browsing the internet; going on trips or engaging in sports; storytelling; conversing; organizing events (e.g., birthday parties); assisting with tasks (e.g., preparing a backpack, tidying belongings); supervising or waiting for the child; accompanying the child (e.g., to the doctor); helping with homework; communicating with teachers or other adults in official roles (for school or extracurricular activities); and providing medical care or transportation to medical appointments. The difference is weighted by the male time spent on the same activities.

### 3.1.1.4 Other Relevant Variables

I collect through the questionnaire couples' bargaining power and individual social norms, used for sub-populations analysis in Section.

Individual Bargaining Power: Based on the question, "In the couple, who usually makes economic decisions (for example, related to financial investments or buying expensive goods)?". I construct a variable coded as 1 if the respondent makes decisions alone or jointly with the partner, and 0 if the partner makes decisions alone.

Individual Gender Norms: An aggregate index (calculated as the mean response for each item) based on agreement (0–100 scale) with the following statements:

- "The task of a man is to contribute to the family income, and the task of a woman is to take care of the children."

- "A preschool-age child (0–6 years) suffers when their mother works."

- "A school-age child (7–11 years) suffers when their mother works."

- "It is a duty towards society to have children."

- "Both parents should be ready to reduce the time dedicated to work for family reasons."

- "A man must be ready to scale down his personal aspirations for the sake of children and the family."

- "Both the father and the mother should stay at home from work for a few months after the birth of their child."

- "When the woman earns more than the man, tensions may arise in the couple."

- "When the man primarily takes care of the house and children, tensions may arise in the couple."

- "A woman must be ready to scale down her personal aspirations for the sake of children and the family."

Individual Parenthood Norms: Measured through exposure to vignettes depicting stereotypes about fatherhood and motherhood, followed by agreement (0–100 scale) with the statement: "I would describe as in the vignettes the dads and moms depicted." Figure 1 presents the vignettes administered to participants.

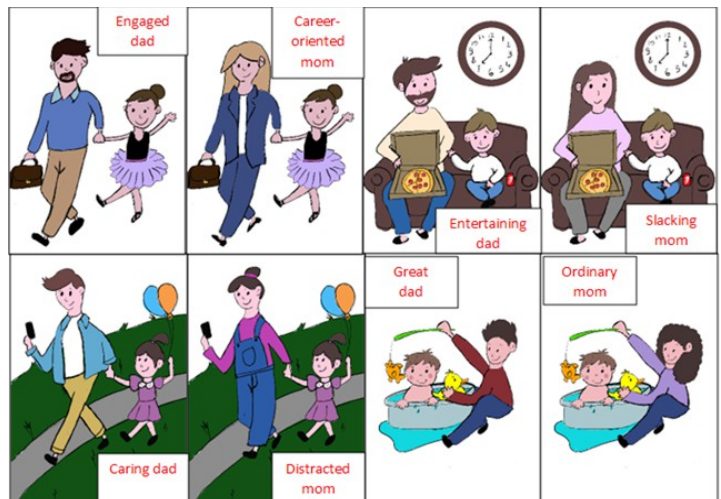

*Figure 1: Vignettes on parenthood norms*

Understanding attitudes toward domestic violence requires capturing both direct judgments about specific behaviors and the broader normative frameworks that shape how individuals interpret gender roles and power dynamics within intimate relationships. The vignette-based indicators allow for the measurement of respondents' immediate reactions to realistic scenarios, including their assessment of the seriousness of the behavior, attribution of responsibility, and degree of victim-blaming—dimensions that are central to how domestic violence is perceived, tolerated, or condemned in society.

Meanwhile, measures of gender norms, masculinity ideals, and parenthood expectations help contextualize these attitudes within the cultural beliefs that legitimize or challenge control, dominance, and traditional role divisions. For instance, endorsing beliefs that men should be the primary breadwinners or that working mothers harm children may correlate with a higher tolerance for controlling or abusive behavior. Attitudes toward domestic violence may be distributed along a continuum shaped by both gender norms and contextual triggers as prior research in economic psychology and development studies demonstrates. Reitmanna et al. (2020), in a survey experiment conducted in Tunisia, found that priming respondents with information about the prevalence of domestic violence significantly reduced its acceptability among men, while framing questions around equality further dampened tolerance among both genders. These effects suggest that cultural norms surrounding violence are not immutable, but rather malleable in response to framing and information cues. Importantly, these dynamics may be intensified or mitigated by the degree of economic autonomy within couples.

Similarly, understanding domestic violence requires not only measuring individual attitudes but also situating them within broader household structures and economic arrangements. Financial management within couples is rarely neutral or purely pragmatic. Rather, control over money often reflects gendered power asymmetries that can shape relationship dynamics and, in some contexts, exacerbate vulnerability to intimate partner violence (Pahl, 1989; Kirchler, 1995). Male-controlled financial systems tend to concentrate discretionary decision-making and personal spending in men's hands, particularly in higher-income households, while female-controlled systems are typically more constrained and associated with budgeting for family necessities (Pahl, 1989). These patterns may contribute to household environments where financial dependence or exclusion amplifies the risk of coercive behaviors and normative justifications of violence.

In this light, the correlation between unequal financial management systems and attitudes justifying violence can be interpreted through the lens of constrained autonomy. When individuals lack control over their own time or money—two fundamental resources—the space for negotiation narrows, and justifications for control may become normalized. This is particularly relevant in couples where traditional gender roles are internalized or where masculinity is closely tied to authority (Meier-Pesti Penz, 2008).

The inclusion of both time diaries and vignette-based attitudinal items makes it possible to go beyond stated preferences and observe lived experience, a methodological advancement echoed by Kirchler (1995) and Pahl (1989). The TIMES dataset provides a rare opportunity to investigate interrelated mechanisms within households. By capturing time-use behaviors alongside attitudinal measures from both partners, it allows for an empirical examination of whether economic decision-making—such as control over household resources or child-rearing time—correlates with higher tolerance for domestic violence or with more traditional views of gender roles. Such an approach aligns with calls for a deeper behavioral analysis of power in the domestic sphere.

Thus, alongside gender gaps in time devoted to unpaid work, I also construct, from time-use diaries, two behavioral outcome variables used in Section 6.

Time spent in leisure with partner: Total weekly hours spent in the presence of one's partner doing activities classified as leisure. These include reading, using social media, watching TV or movies, listening to music or podcasts, exercising, engaging in creative activities, browsing the internet, playing video games, gardening, socializing (e.g., visiting friends or family, dining out, attending events), as well as time spent sleeping and on personal care. This definition is a broader version of that used by (Agüiar and Hurst, 2007).

Time spent in leisure with partner and children: Total weekly hours spent jointly with both partner and children engaging in the above-defined leisure activities.

Only primary activities are considered in constructing these variables.

Figure 2 shows the distribution of weekly hours that participants reported spending leisure time with their partner and with their children.

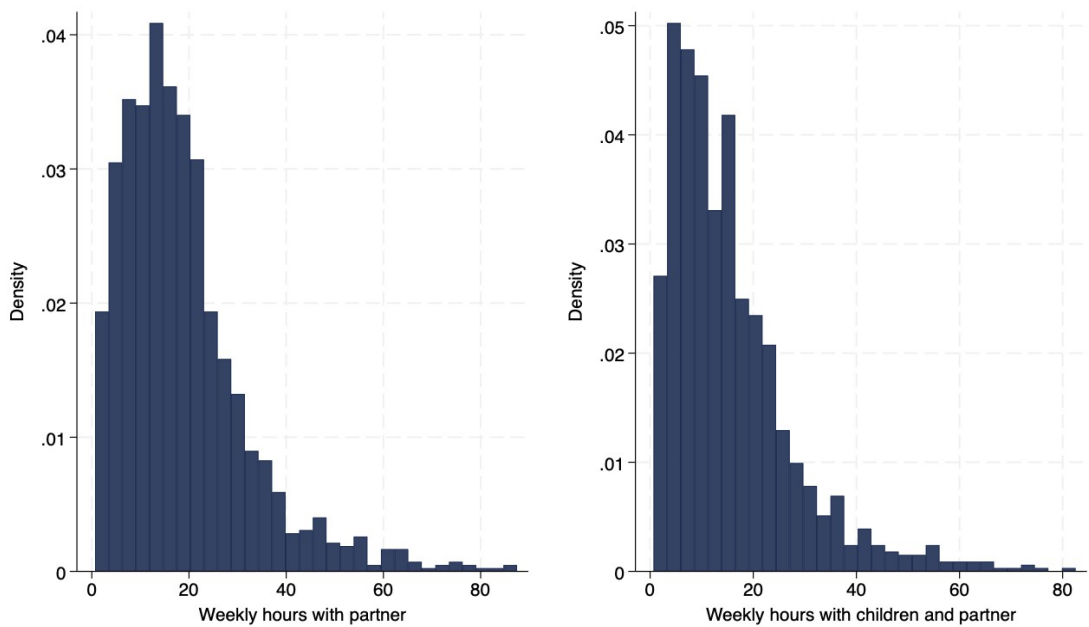

*Figure 2: Distribution of time-use outcomes (weekly hours)*

Furthermore, I compute the relative difference in leisure time—spent with the partner and jointly with the partner and children—declared by men and women within the same couple, measured as a proportion of their average leisure time.

This scales the difference relative to the average time, making the result comparable across couples regardless of how much total time they spend together.

Given that the data include matched reports from both partners in a couple, calculating within-couple differences ensures that the measure reflects directly comparable observations under shared conditions. This approach improves measurement consistency by reducing the influence of unobserved heterogeneity across households (e.g., differences in total available time, employment status, or family structure). It allows for a more accurate assessment of gender asymmetries in reported time use than between-group comparisons, which may conflate structural differences with gender specific reporting.

A value of 0 means that both partners report the same amount of leisure time spent together. A positive value means that the female partner reports more leisure time with the partner than the male partner does. A negative value means that the male partner reports more leisure time with the partner than the female partner does.

Both for time spent together as a couple and for time spent jointly with children, the average difference is negative, indicating that men report spending more time in shared leisure than women do. Specifically, women report approximately 10% less leisure time with their partner—and with both partner and children—compared to men's reports. This discrepancy may reflect differences in perception, where women are less likely to classify certain joint moments as leisure, or it may signal unequal participation in relational or emotional labor within the household.

Together, the variables above aim to capture meaningful aspects of household emotional and relational life that are often gendered. Time spent in leisure together reflects opportunities for shared enjoyment, emotional connection, and informal communication—dimensions of caregiving and relationship maintenance frequently shaped by gender norms and expectations. In particular, the distribution and quality of joint leisure time may reflect internalized beliefs about traditional gender roles, hypermasculinity, and attitudes toward domestic violence, all of which influence who is expected to provide emotional support or participate in relational labor within the household. To the best of my knowledge, this is the first paper to jointly collect data on domestic violence, masculinity, and time-use diaries.

## 3.2 Composite Indicator Construction

### 3.2.1 Dimensionality and Structural Equation Modeling

To examine the relationships among individual attitudes toward intimate partner violence, gender norms, and time use in the household, I conducted an Exploratory Factor Analysis to assess the dimensionality of the observed items. To determine the number of factors to retain, I relied on parallel analysis, which compares the eigenvalues from the actual data with those obtained from randomly generated datasets. Figure 3 shows the results of the parallel analysis.

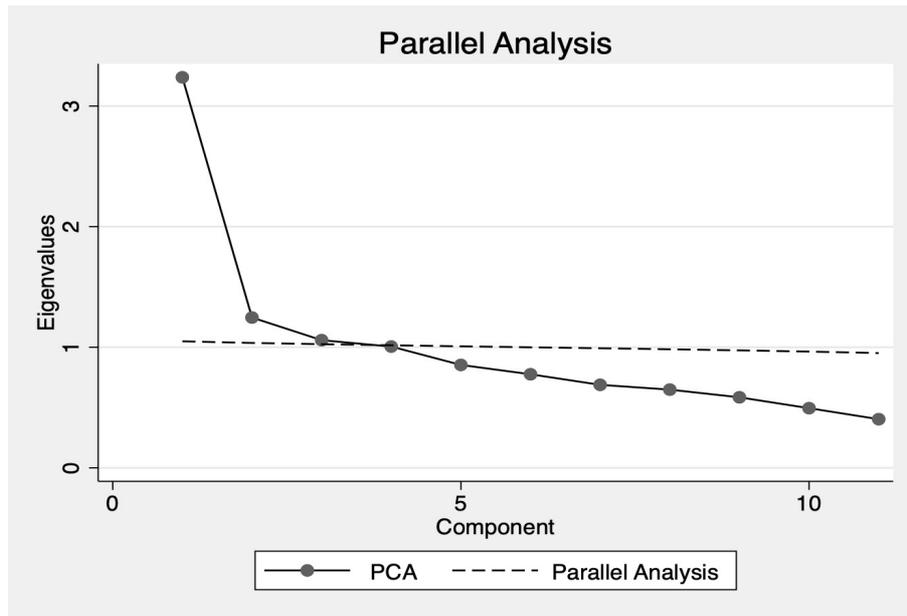

*Figure 3: Parallel analysis of eigenvalues from principal component analysis*

The solid line represents the eigenvalues from the principal component analysis (PCA) of the observed data, while the dashed line corresponds to the 95th percentile eigenvalues from randomly generated data. According to the standard criterion, only those components with eigenvalues greater than the corresponding random eigenvalues should be retained.

As shown in the plot, only the first three components have eigenvalues that exceed the threshold set by the parallel analysis. This result supports the presence of three meaningful latent factors in the data, with the third component close to the cutoff (eigenvalue=1). This finding provides empirical justification for modeling three separate but potentially correlated latent constructs, rather than collapsing them into a single scale. Based on this, I retain three factors and proceed with a confirmatory Structural Equation Model (SEM) with three latent variables: Justification of domestic violence, Masculinity, and Gender gap in unpaid work.

Each latent construct is measured using multiple observed indicators, which are specified as reflective indicators:

$$Seriousness\ of\ Violence = \lambda_1 \cdot Justification + \varepsilon_1$$

$$Victim\ Blaming = \lambda_2 \cdot Justification + \varepsilon_2$$

$$Perpetrator\ Accountability = \lambda_3 \cdot Justification + \varepsilon_3$$

$$Justification\ of\ Domestic\ Violence = \lambda_4 \cdot Justification + \varepsilon_4$$

$$Emotional\ strength = \lambda_5 \cdot Masculinity + \varepsilon_5$$

$$Drinking = \lambda_6 \cdot Masculinity + \varepsilon_6$$

$$Minimization\ of\ harassment = \lambda_7 \cdot Masculinity + \varepsilon_7$$

$$Physical\ strength = \lambda_8 \cdot Masculinity + \varepsilon_8$$

$$Emotional\ toughness = \lambda_9 \cdot Masculinity + \varepsilon_9$$

$$Gender\ gap\ household\ chores = \lambda_{10} \cdot Gender\ gap\ unpaid\ work + \varepsilon_{10}$$

$$Gender\ gap\ childcare = \lambda_{11} \cdot Gender\ gap\ unpaid\ work + \varepsilon_{11}$$

Each indicator is modeled as a linear function of a single latent construct plus a measurement error term εi. All latent variables are standardized to have mean zero and unit variance. Factor loadings (λi) are estimated freely, with one loading per construct fixed to 1 for identification.

The model does not impose directional structural paths among latent variables. Instead, it specifies covariances among the latent constructs to capture their interrelationships. These covariances reflect the hypothesis that the justification of domestic violence, adherence to masculine norms, and household time use are jointly shaped by shared underlying sociocultural factors.

I estimate the model using maximum likelihood with robust standard errors. The model estimated through structural equation modeling (SEM) validates the presence of three distinct latent constructs—Justification of domestic violence, Masculinity, and Gender gap in unpaid work—each captured by multiple reflective indicators. Table 1 presents the results of the estimated model.

*Table 1: Model estimate*

| Observed Variable | Latent Factor | Unstd. Coeff. | Std. Coeff. | Std. Err. | z | p-value |
|---|---|---|---|---|---|---|
| Physical strength | Masculinity | 13.746 | .423 | .329 | 42.16 | <0.001 |
| Emotional strength | Masculinity | 15.503 | .677 | .227 | 68.27 | <0.001 |
| Emotional toughness | Masculinity | 15.49 | .690 | .216 | 71.54 | <0.001 |
| Minimization of harassment | Masculinity | 12.54 | .462 | .285 | 43.93 | <0.001 |
| Drinking | Masculinity | 14.31 | .498 | .288 | 49.59 | <0.001 |
| Justification of Domestic Violence | Justification | 8.82 | .416 | .228 | 38.63 | <0.001 |
| Perpetrator Accountability | Justification | 15.05 | .667 | .249 | 60.27 | <0.001 |
| Victim Blaming | Justification | 13.99 | .553 | .263 | 53.10 | <0.001 |
| Seriousness of Violence | Justification | 13.33 | .601 | .243 | 54.70 | <0.001 |
| Gender gap in household chores | Gender gap in unpaid work | 1.02 | .112 | .323 | 3.16 | 0.002 |
| Gender gap in childcare | Gender gap in unpaid work | 1.14 | .282 | .346 | 3.31 | 0.001 |

All factor loadings are statistically significant (p < 0.001), with standardized coefficients ranging from approximately 0.11 to 0.69. These values suggest moderate to strong associations between the observed variables and their respective latent constructs, confirming the appropriateness of the measurement structure.

Model fit is assessed using standard SEM diagnostics. Table 2 provides equation level goodness of fit statistics, showing that the R-squared values for most indicators fall within an acceptable range.

*Table 2: Equation-Level Goodness of Fit Statistics*

| Observed Variables | Variance | | | R-squared | mc | mc2 |
|---|---|---|---|---|---|---|
| | Fitted | Predicted | Residual | | | |
| Physical strength | 1051.792 | 188.9756 | 862.8163 | 0.1796701 | 0.4238751 | 0.1796701 |
| Emotional strength | 523.1136 | 240.3733 | 282.7404 | 0.4595049 | 0.6778679 | 0.4595049 |
| Emotional toughness | 503.7534 | 240.0157 | 263.7377 | 0.4764548 | 0.6902571 | 0.4764548 |
| Minimization of harassment | 735.2562 | 157.4338 | 577.8164 | 0.2141228 | 0.4627343 | 0.2141228 |
| Drinking | 823.4362 | 204.8036 | 618.6326 | 0.2487182 | 0.4987164 | 0.2487182 |
| Justification of Domestic Violence | 448.8415 | 77.9037 | 370.9378 | 0.1735662 | 0.4166127 | 0.1735662 |
| Perpetrator Accountability | 509.0743 | 226.7054 | 282.3679 | 0.4453296 | 0.6673302 | 0.4453296 |
| Victim Blaming | 639.9978 | 195.8456 | 444.1522 | 0.3060098 | 0.5531815 | 0.3060098 |
| Seriousness of Violence | 491.1438 | 177.9263 | 313.2175 | 0.3626292 | 0.6018838 | 0.3626292 |
| Gender gap in household chores | 83.37905 | 1.046148 | 82.3329 | 0.0125469 | 0.1120129 | 0.0125469 |
| Gender gap in childcare | 16.42664 | 1.313552 | 15.11309 | 0.0799647 | 0.2827804 | 0.0799647 |
| Overall | | | | 0.8831087 | | |

Note. mc = correlation between the observed variable and its predicted value (model correlation). mc2 = squared multiple correlation (Bentler–Raykov coefficient), equivalent to R2 in this model.

The indicators for Masculinity and Justification of domestic violence account for approximately 17% - 48% of the variance in their respective observed variables. In contrast, the indicators for Gender gap in unpaid work show lower R-squared values (0.013 and 0.080), reflecting weaker—but still statistically significant—loadings, which is expected given their more behavioral nature. The overall coefficient of determination for the model is high (CD = 0.883), and the SRMR is 0.057, below the conventional 0.08 threshold, indicating good model fit.

Table 3 presents the structural part of the model, capturing the covariance relationships among the three latent factors, with masculinity norms shaping both the other latent constructs.

$$Tolerant\ Attitudes\ Towards\ Domestic\ Violence_i = \phi_1 \cdot z(Justification_i) + \phi_2 \cdot z(Masculinity_i) + \phi_3 \cdot z(Time_i) rev$$

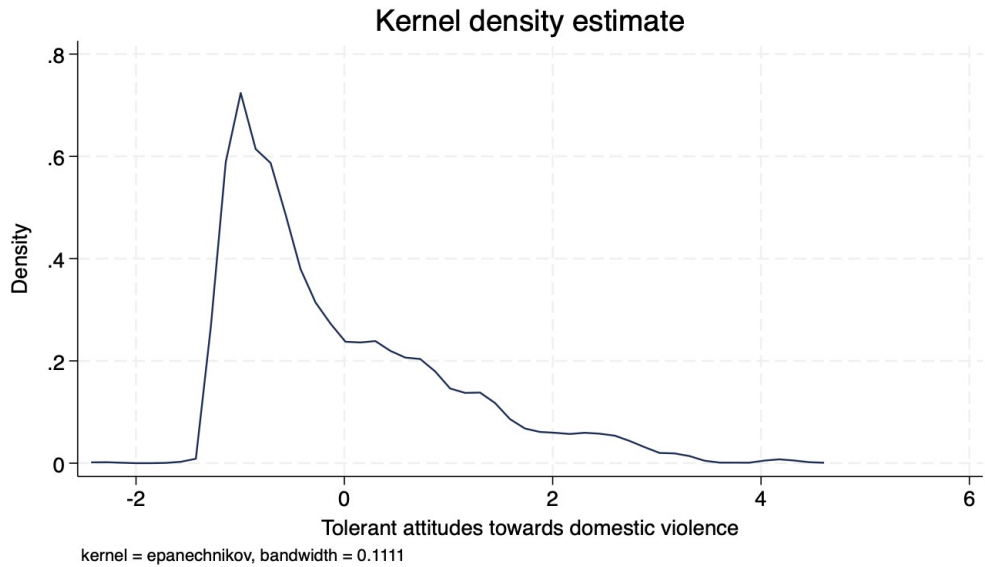

*Figure 4: Distribution of the proposed measure of tolerant attitudes towards domestic violence*
*Table 3: Structural Relationships Among Latent Variables and Reliability of Composite Indicator*

| Latent 1 | Latent 2 | Relation | Estimate | Std. Err. | z |
|---|---|---|---|---|---|
| Masculinity | Justification of domestic violence | Covariance | .799 | .009 | 81.27 |
| Masculinity | Gender difference in unpaid work | Covariance | -.170 | .052 | -3.26 |
| Reliability of Composite Indicator | | | | | |
| Average interitem covariance | | | .263 | | |
| Number of items in the scale | | | 3 | | |
| Scale reliability coefficient (Cronbach's alpha) | | | .777 | | |

The strongest correlation is between Masculinity and Justification of domestic violence (r = 0.799, p < 0.001), suggesting that stronger masculine norms are closely associated with the justification of intimate partner violence. A moderate negative covariance is observed between Masculinity and Gender difference in unpaid work (r = –0.170, p = 0.001). The third latent construct, Gender difference in unpaid work, captures gender asymmetries in the division of household labor. Higher values of Gender differences in unpaid work indicate greater female overrepresentation in unpaid care work, interpreted as normative patterns of gendered time use. The negative relationship, thus, implies that stronger endorsement of masculine norms is associated with greater acceptance of unequal gender roles in domestic responsibilities.

Although the loadings for the Gender difference in unpaid work indicators are statistically significant, they are smaller than those for the other constructs (standardized loadings ≈ 0.11), and the corresponding R-squared values are low (0.013 and 0.080). This is expected, as time-use asymmetries may reflect both normative beliefs and practical constraints. Still, a reliability analysis of the three indicators yields a Cronbach's alpha of 0.777 and an average inter-item covariance of 0.263, supporting the internal coherence of the scale used to construct the composite indicator Attitude towards domestic violence (Table 3).

### 3.2.2 Composite Indicator Construction

Based on the estimated latent constructs, I then construct a composite indicator, Tolerant Attitudes Towards Domestic Violence, which summarizes individual positioning along the three dimensions captured by the SEM.

The composite indicator is computed as the first principal component from a Principal Component Analysis (PCA) of the three standardized variables. This strategy combines the dimensions into a single measure that captures the dominant pattern of co-variation. In other words, it summarizes how aligned individuals are with a broader attitudinal profile that includes tolerance of violence, gendered norms, and household inequality. Formally, for individual i, the indicator is defined as:

where $z(\cdot)$ denotes the standardized score, $z(Time_i)rev$ is the reverse-coded time use factor, and $\phi_1$, $\phi_2$, $\phi_3$ are the loadings from the first principal component. These loadings are chosen to maximize the variance explained by the linear combination and ensure that the composite indicator captures the dominant shared variance across the three dimensions.

This composite indicator is used as the proposed summary measure of individual level tolerant attitudes towards domestic violence. Figure 4 shows the kernel density estimate of the resulting composite indicator.

The distribution is right-skewed, with a sharp mode just below zero, indicating that the majority of respondents express non-tolerant attitudes toward domestic violence. Nonetheless, the long right tail reveals a non-trivial share of individuals who display more tolerant or permissive views. This asymmetry is informative: it suggests that while most respondents reject violence-supportive norms, a meaningful minority holds ambivalent or accepting positions. Such variation underscores the importance of modeling attitudes on a continuous scale, rather than using a binary classification, to capture both outright rejection and partial endorsement of domestic violence within prevailing gender norms.

## 4. Results

The empirical analysis validates the proposed composite index of tolerant attitudes toward domestic violence and demonstrates its relevance for household behavior, family dynamics, and demographic outcomes. The results are presented in three steps: (i) validation with individual characteristics, (ii) validation with partner characteristics, and (iii) associations with household time allocation.

## 4.1 Individual-level Correlates

Table 4 shows the associations between the composite measure and respondents' characteristics.

*Table 4: Validation of the proposed measure against individual characteristics*

|                                          | (1)       | (2)       | (3)      | (4)      | (5)       | (6)       |
|------------------------------------------|-----------|-----------|----------|----------|-----------|-----------|
| Female                                   | –0.196*   |           |          |          |           |           |
|                                          | (0.0550)  |           |          |          |           |           |
| Individual Level of Education            |           | –0.0436*  |          |          |           |           |
|                                          |           | (0.0162)  |          |          |           |           |
| Employed = 1                             |           |           | –0.0880  |          |           |           |
|                                          |           |           | (0.0562) |          |           |           |
| Individual Bargaining Power = 1          |           |           |          | –0.172*  |           |           |
|                                          |           |           |          | (0.103)  |           |           |
| Individual Conservative Gender Norms     |           |           |          |          | 0.0151*   |           |
|                                          |           |           |          |          | (0.00105) |           |
| Individual Conservative Parenthood Norms |           |           |          |          |           | 0.0131*   |
|                                          |           |           |          |          |           | (0.000802)|
| Constant                                 | 0.129*    | 0.287*    | 0.0680   | 0.166*   | –0.586*** | –0.350*** |
|                                          | (0.0452)  | (0.109)   | (0.0466) | (0.0996) | (0.0427)  | (0.0286)  |
| Observations                             | 1,730     | 1,730     | 1,730    | 1,730    | 1,730     | 1,730     |
| R-squared                                | 0.008     | 0.005     | 0.001    | 0.002    | 0.094     | 0.142     |

Standard errors in parentheses
*** $p<0.01$, ** $p<0.05$, * $p<0.1$

The coefficients show that socio-demographic variables such as gender, education, employment, or reported bargaining power are not significantly related to tolerance of domestic violence. In contrast, conservative gender and parenthood norms are highly significant predictors: a one-point increase in either scale corresponds to an increase of approximately 0.02 points in the tolerance index ($p<0.01$).

These results underscore that cultural orientations, rather than demographic characteristics, drive tolerance of violence. The magnitude of the coefficients is modest but highly robust, suggesting that small shifts in norms at the individual level may accumulate into meaningful household-level effects. This highlights the centrality of normative environments in shaping household behavior, fertility choices, and women's labor supply, beyond what can be explained by income or education alone.

## 4.2 Partner-level Correlates. Women only

Table 5 examines women's attitudes in relation to their partners' characteristics.

*Table 5: Validation of the proposed measure against women's partner characteristics*

|  | (1) | (2) | (3) | (4) | (5) | (6) |
|---|---|---|---|---|---|---|
| Partner Level of Education | –0.0156 |  |  |  |  |  |
|  | (0.0350) |  |  |  |  |  |
| Partner Employed = 1 |  | –0.424* |  |  |  |  |
|  |  | (0.131) |  |  |  |  |
| Partner Bargaining Power = 1 |  |  | –0.184 |  |  |  |
|  |  |  | (0.225) |  |  |  |
| Assigns Caregiving to Women = 1 |  |  |  | –0.147 |  |  |
|  |  |  |  | (0.109) |  |  |
| Partner Conservative Gender Norms |  |  |  |  | 0.0107* |  |
|  |  |  |  |  | (0.00213) |  |
| Partner Conservative Parenthood Norms |  |  |  |  |  | 0.0111* |
|  |  |  |  |  |  | (0.00172) |
| Constant | 0.0372 | 0.284 | 0.135 | 0.00951 | –0.428* | –0.308* |
|  | (0.165) | (0.120) | (0.219) | (0.0604) | (0.0859) | (0.0553) |
| Observations | 454 | 454 | 454 | 454 | 454 | 454 |
| R-squared | 0.001 | 0.029 | 0.002 | 0.004 | 0.046 | 0.091 |

Standard errors in parentheses.
*** $p<0.01$, ** $p<0.05$, * $p<0.1$

Several findings stand out. First, male partner education is negatively associated with women's tolerance (–0.086, $p<0.05$), consistent with the view that better-educated men foster more egalitarian household environments. Second, when men assign caregiving exclusively to women, female respondents score substantially lower on the tolerance index (-0.33, $p<0.05$), suggesting that structural inequalities in caregiving reduce women's internalization of norms legitimizing control. Third, partners' conservative gender and parenthood norms are both strongly associated with women's tolerance (coefficients ≈0.015–0.021, p<0.01).

These results indicate that women's tolerance is shaped not only by their own attitudes but also by their partners' normative environment and household role assignments. This finding is consistent with bargaining models in which intra-household inequality constrains women's agency, leading to a rationalization of coercive dynamics. From a demographic perspective, this points to mechanisms of intergenerational transmission of norms: when women normalize unequal caregiving and tolerate coercion, children are more likely to inherit these attitudes.

## 4.3 Household Time Allocation

A distinctive feature of this study is the link between tolerance of violence and household time-use patterns. Table 6 presents regressions of the composite index on time spent in leisure with partners and children.

*Table 6: Validation of the proposed measure against household time allocation*

|  | (1) | (2) |
|---|---|---|
|  | Minutes Leisure with Partner | Minutes Leisure with Partner and Children |
| Tolerant Attitudes (Index) | 0.783** | 1.335*** |
|  | (0.309) | (0.306) |
| Constant | 18.70*** | 15.50*** |
|  | (0.336) | (0.338) |
| Observations | 1,511 | 1,259 |
| R-squared | 0.004 | 0.014 |

Standard errors in parentheses.
*** p<0.01, ** p<0.05, * p<0.1

Higher tolerance scores are positively associated with reported leisure time, corresponding to an increase of approximately 0.78 minutes per unit of tolerance when spending time with the partner and 1.34 minutes when spending time jointly with both partner and children ($p < 0.05$ and $p < 0.01$, respectively).

Although the effect sizes are modest, the direction of association remains positive and statistically significant, indicating that higher tolerance of domestic violence corresponds to slightly greater reported shared leisure time. This result is somewhat counterintuitive: while tolerance of violence is generally associated with lower relationship quality, it appears here to coexist with more frequent co-presence.

Several mechanisms may explain this pattern: (i) male authority in structuring leisure, where joint time reflects male preferences rather than equal participation; (ii) women's rationalization of co-presence, whereby activities under male

|  | Female | Male | Low Edu | High Edu | Not Empl. | Empl. | Low Barg. | High Barg. | Non-Conserv. | Conserv. | No Charity | Charity | No Center | Center | Way Out | No Way Out |
|---|---|---|---|---|---|---|---|---|---|---|---|---|---|---|---|---|
| Tolerant attitudes | 1.166*** | 1.473*** | 1.156*** | 1.621*** | 0.948* | 1.524*** | 2.742*** | 1.186*** | 1.318*** | 1.164** | 1.641*** | 0.761 | 1.435*** | 1.056 | 0.655 | 1.667*** |
|  | (0.417) | (0.455) | (0.372) | (0.521) | (0.531) | (0.376) | (0.788) | (0.329) | (0.384) | (0.588) | (0.384) | (0) | (0.359) | (0.646) | (0.542) | (0.375) |
| Constant | 15.70*** | 15.40*** | 15.79*** | 15.03*** | 15.38*** | 15.57*** | 12.99*** | 15.71*** | 15.39*** | 15.96*** | 15.77*** | 14.89 | 15.55*** | 15.46*** | 15.36*** | 15.61*** |
|  | (0.528) | (0.446) | (0.451) | (0.495) | (0.594) | (0.412) | (1.001) | (0.354) | (0.391) | (0.807) | (0.413) | (0) | (0.371) | (0.932) | (0.606) | (0.407) |
| Obs. | 497 | 762 | 778 | 481 | 414 | 845 | 113 | 1,146 | 962 | 297 | 876 | 383 | 1,052 | 207 | 309 | 950 |
| R-squared | 0.014 | 0.014 | 0.010 | 0.025 | 0.007 | 0.018 | 0.077 | 0.011 | 0.012 | 0.010 | 0.019 | 0.006 | 0.015 | 0.010 | 0.005 | 0.018 |

control are perceived as shared leisure; and (iii) conflation of quantity and quality, since more time together does not necessarily indicate higher relational quality.

From an economic perspective, these findings are important because time-use data are often interpreted as indicators of family cohesion. The evidence here suggests that co-presence may conceal rather than reveal equality—highlighting the need to interpret shared time in light of underlying gender norms and bargaining dynamics within the household.

## 4.4 Subgroup Analyses

I estimate separate regressions within the subgroups. The subgroups are constructed from individual characteristics and survey responses, and they reflect both demographic and attitudinal heterogeneity:

1. Gender: Individuals are split into two groups based on self-reported gender (men vs. women).

2. Education: Respondents are classified as above or below the sample median of years of education.

3. Employment Status: Individuals are grouped based on whether they report being currently employed or not.

4. Bargaining Power: This indicator equals 1 if the individual reports having a say in major household financial decisions and 0 otherwise.

5. Conservative Parenthood Norms: Based on agreement with vignette statements suggesting gender-typical parenting roles. Individuals above the median score are classified as holding more conservative views.

6. Domestic Violence Charity Intention: Derived from the question: "I am inclined to donate my compensation for participating in the survey to an association that supports victims of violence against women/men" (0 = not at all, 100 = completely). Individuals above the median are coded as more supportive of domestic violence-related causes.

7. Knowledge of Domestic Violence Centers: Based on the item: "Is there a gender violence center close to where you live whose activities you are familiar with?" (0 = not familiar at all, 100 = perfectly familiar). Individuals above the median are categorized as more aware of domestic violence resources.

8. Belief in a Way Out of domestic violence: From the question: "There is a way out of violence against women" (0 = never, 100 = always). This measure captures perceived agency in escaping domestic violence. Individuals above the median are coded as having greater belief in the possibility of change.

9. Each subgroup is defined as a binary variable using the sample median as a threshold. This approach ensures sufficient balance between groups and allows for interpretable comparison of regression coefficients across clusters.

Tables 7 and 8 present subgroup regressions.

*Table 7: Validation of the proposed measure against household time allocation (time with partner). Subgroup analyses.*

| | Female | Male | Low Edu | High Edu | Not Empl. | Empl. | Low Barg. | High Barg. | Non-Conserv. | Conserv. | No Charity | Charity | No Center | Center | Way Out | No Way Out |
|---|---|---|---|---|---|---|---|---|---|---|---|---|---|---|---|---|
| Tolerant attitudes | 0.677 | 0.773* | 0.402 | 1.432*** | 0.328 | 0.991*** | 2.716*** | 0.550* | 0.671* | 0.391 | 0.746** | 0.889 | 0.774** | 0.557 | 0.305 | 1.100*** |
| | (0.436) | (0.444) | (0.371) | (0.532) | (0.559) | (0.371) | (0.920) | (0.324) | (0.383) | (0.601) | (0.380) | (—) | (0.365) | (0.642) | (0.536) | (0.382) |
| Constant | 19.45*** | 18.19*** | 18.94*** | 18.29*** | 18.94*** | 18.59*** | 17.41*** | 18.77*** | 18.31*** | 20.07*** | 19.09*** | 17.75 | 18.56*** | 19.49*** | 17.68*** | 19.09*** |
| | (0.542) | (0.433) | (0.439) | (0.514) | (0.622) | (0.400) | (1.181) | (0.347) | (0.391) | (0.771) | (0.400) | (—) | (0.370) | (0.936) | (0.599) | (0.407) |
| Obs. | 609 | 902 | 939 | 572 | 485 | 1,026 | 129 | 1,382 | 1,152 | 359 | 1,063 | 448 | 1,264 | 247 | 380 | 1,131 |
| R-squared | 0.004 | 0.003 | 0.001 | 0.015 | 0.001 | 0.007 | 0.048 | 0.002 | 0.003 | 0.001 | 0.004 | 0.006 | 0.004 | 0.002 | 0.001 | 0.007 |

Notes: Standard errors clustered at the couple level in parentheses. Subgroups: education, employment status, bargaining power, parenthood norms, intimate partner violence-related awareness/engagement. *** p<0.01, ** p<0.05, * p<0.1

*Table 8: Validation of the proposed measure against household time allocation (time with partner and children). Subgroup analyses.*

Notes: Standard errors clustered at the couple level in parentheses. Subgroups: education, employment status, bargaining power, parenthood norms, intimate partner violence-related awareness/engagement. *** p<0.01, ** p<0.05, * p<0.1

The positive association between tolerance and leisure persists across most groups, though its magnitude and significance vary. Effects are: (i) stronger for men than women, suggesting that men's reports of shared leisure are more influenced by tolerant attitudes; (ii) amplified among individuals aware of domestic violence services or believing in a way out, which may reflect greater recognition of violence without necessarily altering household practices; (iii) weaker or absent among individuals with limited perceived agency, highlighting the interaction between cultural norms and perceived alternatives.

These patterns reinforce the idea that attitudes and behaviors are interconnected but mediated by awareness and gendered perceptions.

Taken together, the findings provide three main insights: (i) norms outweigh demographics: cultural orientations are the strongest predictors of tolerance, not education, gender, or employment; (ii) partner characteristics matter: women's tolerance is shaped by men's education, caregiving role assignments, and normative orientations; (iii) tolerance is linked to more joint leisure: counterintuitively, tolerant households report greater co-presence, suggesting that time together can coexist with hierarchical dynamics.

These findings underscore that tolerance of violence is not just an attitudinal marker but a structural feature of household organization. It influences fertility choices, labor supply, and the intergenerational transmission of gender norms. Moreover, it complicates the interpretation of time-use data, showing that shared time may reinforce rather than challenge traditional authority structures.

# 5. Information Treatment

To examine whether individuals' attitudes toward domestic violence are malleable to informational cues, I implemented an information treatment designed to increase the salience of physical and psychological violence.

## 5.1 Experimental Design

The information treatment was designed to test whether making violence more salient and cognitively accessible can shift participants' tolerance toward domestic violence.

The experiment relies on a two-by-two design. Participants were randomly assigned to one of the two following baseline vignettes:

Physical violence vignette:

"*Sara and Davide have been a couple for ten years. During one of their many arguments, Sara started yelling and Davide slapped and hit her.*"

Psychological violence vignette:

"*Sara and Davide have been a couple for ten years. When Sara goes out at night, Davide constantly messages her to ask what she is doing, where she is, and whom she is with.*"

These vignettes capture distinct but complementary dimensions of intimate-partner violence—physical aggression versus controlling behavior.

Immediately afterward, half of the respondents, randomly selected, received an additional block of information intended to increase salience and institutional awareness:

"*According to the World Bank, 35% of women worldwide have experienced partner violence, and 38% of femicides worldwide are committed by the partner (World Bank, 2019).*"

"*Violence appears in various forms, not always easily identifiable: mistreatment is not only physical or sexual but also psychological and economic (Istat, 2018).*"

*"If a mother experiences violence and wants to protect her children, the Juvenile Court has programs to assist her (Juvenile Court, 2015)."*

After reading the vignette—and, for half of the sample, this additional information—participants answered the set of questions that compose the indicator of tolerant attitudes toward domestic violence.

This composite index aggregates responses to items capturing the perceived acceptability of violence and coercion within intimate relationships, standardized on a 0–1 scale, with higher values denoting greater tolerance.

## 5.2 Results

Table 9 reports the average marginal effects (AME) from probit regressions estimated among women.

Model A compares respondents exposed to the physical-violence vignette versus the psychological-control vignette, while Model B adds the informational treatment described above.

*Table 9: Effect of the information treatment on women's tolerant attitudes towards domestic violence.*

|  | Model A: Physical violence vignette | Model B: Vignette + Information treatment |
|---|---|---|
| Average marginal effect | –0.0523 (0.0296)* | –0.0340 (0.0344) |
| 95% confidence interval | [–0.1104, 0.0057] | [–0.1015, 0.0335] |
| Observations | 1,060 | 1,060 |

The first specification shows that exposure to the physical-violence vignette reduces the probability of justifying violence by approximately 5.2 percentage points (AME = –0.052, SE = 0.029, p = 0.077).

The effect is negative and significant, indicating a modest but meaningful decline in tolerance when violence is presented as overtly physical.

When factual information is provided (Model B), the estimated effect remains negative (AME = –0.034, SE = 0.034, p = 0.32) but is statistically not significant.

Thus, the informational treatment generates a weaker incremental effect than the vignette itself, suggesting that abstract statistics and institutional reassurance do not amplify the emotional impact of the narrative.

The findings reveal two complementary mechanisms. First, salience of visible violence drives attitudinal change: explicit physical aggression elicits stronger moral condemnation and lowers tolerance. Second, informational cues—though factually rich and normatively unambiguous—have limited influence on immediate beliefs.

Providing factual and institutional context may trigger cognitive rationalization ("violence is common" or "institutions can help"), reducing the emotional shock induced by the vignette. This asymmetry highlights a distinction between emotional salience and cognitive information. The former acts as a moral cue that temporarily reduces justification; the latter appeals to reasoning but struggles to alter entrenched social norms. Hence, informational campaigns alone may have limited behavioral traction if not accompanied by emotionally engaging elements that personalize the issue.

From a behavioral-economic perspective, the experiment shows that information alone is not sufficient to reshape tolerance of domestic violence. While exposure to a concrete act (the physical-violence vignette) reduces justification by roughly five percentage points, the subsequent addition of factual and institutional information yields no further improvement.

These findings imply that public awareness initiatives emphasizing statistics or institutional programs may raise knowledge but not transform norms. By contrast, interventions that make violence personally salient and emotionally resonant—for example, storytelling, survivor testimonies, or dramatizations—are more likely to engage empathy and induce moral reevaluation. In practice, information must be emotionally anchored to shift beliefs: purely factual content about prevalence or support services tends to be cognitively processed yet normatively inert.

# 6. Discussion and Conclusion

This paper has introduced a novel framework for measuring attitudes toward domestic violence by integrating survey-based attitudinal data with high-frequency time-use diaries from couples with children. By combining vignette-based measures with behavioral indicators of household inequality, I construct and validate a composite index that captures latent tolerance of domestic violence.

Methodologically, this approach advances the measurement of domestic violence beyond reliance on self-reports. Substantively, it demonstrates that tolerance of violence is closely linked to conservative gender norms, partner characteristics, caregiving asymmetries, and patterns of time allocation. Unexpectedly, tolerant attitudes are associated with more reported joint leisure. This finding challenges conventional assumptions about violence and relational withdrawal, showing that co-presence may coexist with unequal bargaining power and traditional authority.

The implications are threefold. First, tolerance of violence is embedded in household structures that shape fertility decisions, women's labor supply, and child investments. Second, time-use patterns suggest that tolerance affects intergenerational transmission of norms, as children observe both unequal care arrangements and how leisure is structured. Third, the composite index offers policymakers a tool for identifying hidden inequalities in settings where direct reporting is unreliable, with implications for interventions that support family stability and gender equity.

These findings also resonate with recent work that highlights how stereotypes shape perceptions of violence and how deviations from traditional gender roles can provoke violent backlash. My results extend this literature by showing that tolerance is embedded not only in perceptions or income gaps but also in the organization of time within households.

In conclusion, domestic violence should be understood as both a cause and consequence of demographic outcomes. By embedding attitudes in the framework of household allocation, this paper bridges measurement innovation with core themes in economics. Understanding how tolerance of violence interacts with fertility, labor supply, and intergenerational norm transmission is essential for demographic analysis and policy design.